# DETECTA: Investigación de metodologías no intrusivas apoyadas en tecnologías habilitadoras 4.0 para abordar un mantenimiento predictivo y ciberseguro en pymes industriales

A Preprint

Alvaro García*
Fundación Cidaut
Valladolid, Spain
alvgar@cidaut.es

Alejandro Echeverría
Funditec
Madrid, Spain
aecheverria@funditec.es

José Félix Ovejero
Industrias Maxi
Valladolid, Spain
josefelix.ovejero@industrias-maxi.es

June 9, 2023

## Abstract

En esta publicación se presentan los resultados del proyecto DETECTA, que aborda actividades de investigación industrial para la generación de conocimiento predictivo orientado a la detección de anomalías en sistemas de fabricación de piezas mecanizadas. Aborda diferentes retos tecnológicos para mejorar al mismo tiempo la disponibilidad de la maquinaria y la protección frente a ciberamenazas de los sistemas industriales, con la colaboración de centros de conocimiento y expertos en procesos industriales. A través del uso de tecnologías innovadoras como el gemelo digital y la inteligencia artificial implementa metodologías de caracterización de procesos y detección de anomalías de forma no intrusiva sin limitar la productividad de la planta industrial de acuerdo a las necesidades de mantenimiento y acceso remoto. La investigación se ha apoyado en una evaluación general de los entornos conectados en pymes para identificar si los beneficios de la digitalización compensan los riesgos que no se pueden eliminar. Los resultados obtenidos, a través de un proceso de supervisión por expertos de proceso y aprendizaje automático, han permitido discriminar anomalías entre eventos puramente técnicos y eventos relacionados con ciberincidentes o ciberataques.

## 1 Introducción

El constante desarrollo y evolución de las tecnologías habilitadoras de la Industria 4.0 está revolucionando la competitividad de las empresas del sector productivo; en particular por su capacidad para mejorar los diferentes procesos a través de la obtención y gestión de datos en tiempo real. Se estima que sólo con el aprovechamiento de las nuevas tecnologías para la aplicación del mantenimiento predictivo en empresas del sector manufacturero, se puede conseguir una reducción de costes del 21% [Feldman et al., 2017], impactando en prácticamente todas las funciones operativas: producción, logística, inventario y calidad. La Comisión Europea, a través de la iniciativa I4MS (ICT Innovation for Manufacturing SMEs) hace énfasis en que para avanzar hacia la industrialización y la adopción de la tecnología, especialmente en las Pequeñas y Medianas Empresas (PYMEs), es necesario innovar con metodologías que puedan testear su impacto sobre los procesos de fabricación. El propósito es utilizar la información generada en los sistemas ciberfísicos para mejorar sustancialmente, entre otros, la capacitación del personal, la calidad del proceso, la optimización de recursos y la flexibilidad [Carracedo et al., 2019]. A su vez, la digitalización de la industria de la mano de los sistemas ciberfísicos y de la convergencia de las Tecnologías de Operación (TO) y las Tecnologías de la Información (TI), plantea nuevos

---

retos; por un lado, la integración y monitorización de todo tipo de sistemas heterogéneos de muy variadas tecnologías de proceso, y, por otro lado, la mantenibilidad de dichos sistemas en un nuevo paradigma físico-digital que introduce riesgos de ciberseguridad tanto a pie de planta como a lo largo de la comunicación con su cadena suministro. Además, las PYMEs, de manera cada vez más habitual, se enfrentan a entornos heredados no adecuados por diseño para afrontar estos retos [Horváth and Szabó, 2019]. En consecuencia, ha habido un aumento sustancial de los riesgos de seguridad basado en nuevas amenazas específicas, como el acceso remoto, en relación con los sistemas ciberfísicos en redes convergentes TO/TI [Centro Criptológico Nacional de España, 2021]. Circunstancias recientes como la pandemia COVID-19 que desencadenó medidas de limitación de movimientos y de interrupción de cadenas de suministro, han acelerado la digitalización de las operaciones productivas en toda la industria [Czifra and Molnár, 2020]. Por tanto, la necesidad de incorporación de tecnologías en las fábricas que permitan llevar a cabo labores de asistencia de forma remota, se ha puesto a prueba en las funciones de mantenimiento, sostenibilidad y continuidad de los procesos de fabricación, a la vez que se han introducido nuevos riesgos de ciberseguridad.

Frente a ello, las nuevas tecnologías de caracterización físico-digital, como los gemelos digitales [Raptis et al., 2019], permiten representar una abstracción de la realidad de los sistemas industriales sobre múltiples niveles de interacción entre procesos, sistemas y trabajadores dentro de un espacio virtual [Semeraro et al., 2021], por lo que ha despertado el interés de la industria por su contribución a las estrategias de mantenimiento [Errandonea et al., 2020]. Además, el concepto de gemelo digital es capaz de probar la resistencia a ciberataques de infraestructuras compuestas por sistemas de instrumentación y control utilizando técnicas de caja negra o "black box" [Somers et al., 2023] de forma no intrusiva y sin riesgo de daños reales al sistema. Por tanto, la generalización de entornos de monitorización, ampliamente extendidos en la industria, y el aumento de las técnicas de mantenimiento predictivo, establecen las bases para la utilización de gemelos digitales con interfaces hombre-máquina para acciones de mejora continua.

En este trabajo se presenta el proyecto DETECTA, centrado en la generación de conocimiento predictivo mediante la investigación industrial en PYMEs de nuevas metodologías no intrusivas en procesos de fabricación del sector manufacturero. Con la aplicación del concepto de gemelo digital y técnicas de aprendizaje automático, investiga la caracterización de procesos industriales con naturaleza heterogénea, incluyendo la maquinaria involucrada, de manera que se permita una detección de anomalías técnicas de forma temprana. La detección tradicional de anomalías en los procesos productivos se ha basado mayoritariamente en metodologías de aprendizaje reactivas, con alto impacto y costes elevados para los entornos industriales. Sin embargo, la investigación planteada en el proyecto DETECTA, pretende hibridar al mismo tiempo dos enfoques concepto que prestan soporte a los procesos de fabricación; por un lado la productividad asociada al mantenimiento predictivo, y por otro la detección temprana de eventos asociados a las amenazas de ciberseguridad emergentes sobre sistemas industriales conectados. Para responder a estos retos plantea tres soluciones tecnológicas: (i) caracterización del procesos con la disponibilidad de toda la información posible de forma no intrusiva y sin limitar la productividad, (ii) detección temprana de anomalías técnicas, y (iii) discriminación entre causas producidas por eventos técnicos y ciberincidentes.

De acuerdo a este planteamiento, la organización del resto del artículo será la siguiente. En la Sección 2 se incluye un análisis de la situación actual de las tecnologías. En la Sección 3 se describe la metodología de investigación con técnicas de aprendizaje no intrusivas para el diagnóstico predictivo de anomalías de funcionamiento y seguridad. A continuación se abordan los diferentes componentes implementados durante la investigación en la Sección 4. Por último la Sección 5 presenta los resultados y la Sección 6 las conclusiones obtenidas durante el proyecto.

## 2 Revisión de la tecnología

### 2.1 Mantenimiento predictivo y ciberseguro en entornos de fabricación industrial

El diagnóstico de un entorno de fabricación está implícitamente asociado a la identificación del nivel de control y eficiencia sobre sus procesos. La ausencia, en un alto porcentaje, de soluciones de digitalización interconectadas a nivel de planta, supone la principal barrera para el avance hacia la fábrica inteligente y conectada [Kroll et al., 2016]. En ese sentido, la fábrica del futuro se caracteriza por entornos digitalizados, virtuales y sostenibles [Zhong et al., 2017]. A partir de la conectividad, en términos de la disponibilidad de la información y de la capacidad de interacción entre los sistemas de producción, es posible abordar el diagnóstico predictivo [Lee et al., 2013], la visualización y la detección de anomalías [Ibarguengoytia et al., 2016] de una forma más amplia. No obstante, la introducción de tecnologías digitales para acceso remoto a los sistemas de producción, ha aumentado la exposición de los contextos de operación y de ejecución de procesos automatizados, muchos de ellos basados en sistemas heredados con menores exigencias de seguridad o incluso sin la posibilidad de ser actualizados.

A partir de un estudio que incluye el autodiagnóstico entre los miembros de la Asociación Empresarial Innovadora (EAI) del sector de Automoción FACYL[2], se han obtenido diferentes palancas e indicadores relacionados con la madurez y uso de tecnologías digitales desde el punto de vista de la mantenibilidad y ciberseguridad. En la muestra del cuestionario han participado una veintena de empresas del sector: 9 fabricantes de componentes, 7 empresas de mecanizado y 4 ingenierías y empresas de servicios auxiliares.

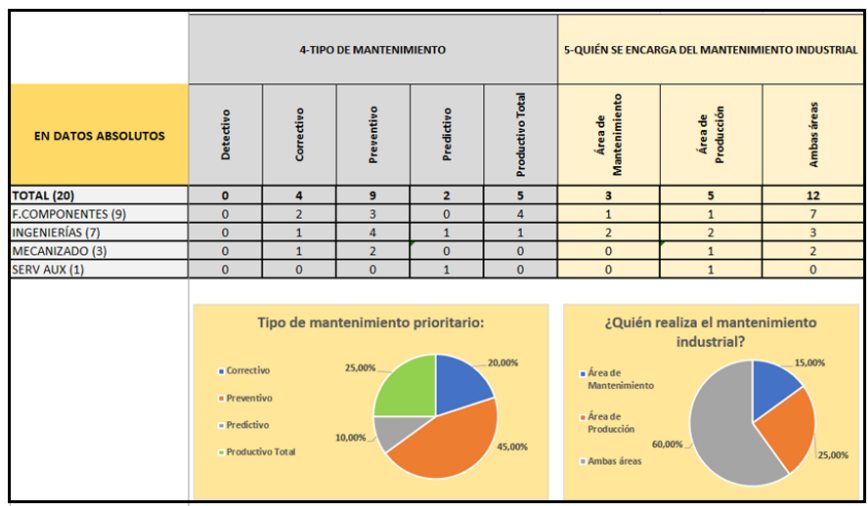

Figure 1: Autodiagnóstico sector Automoción: mantenimiento

En la Figura 1 se observa que todavía el mantenimiento preventivo es la estrategia utilizada en el 45% de los casos, siendo compartido en más de la mitad de los casos entre el departamento de mantenimiento y el área de producción. Esto nos indica que todavía no existe un nivel de especialización de la tecnología suficiente para confiar en técnicas de mantenimiento predictivo, relegadas a apenas el 10% de los casos. Si hablamos de la tecnología del gemelo digital, en general, las PYMEs desconocen las aplicaciones de la implantación de modelos no intrusivos y sus beneficios.

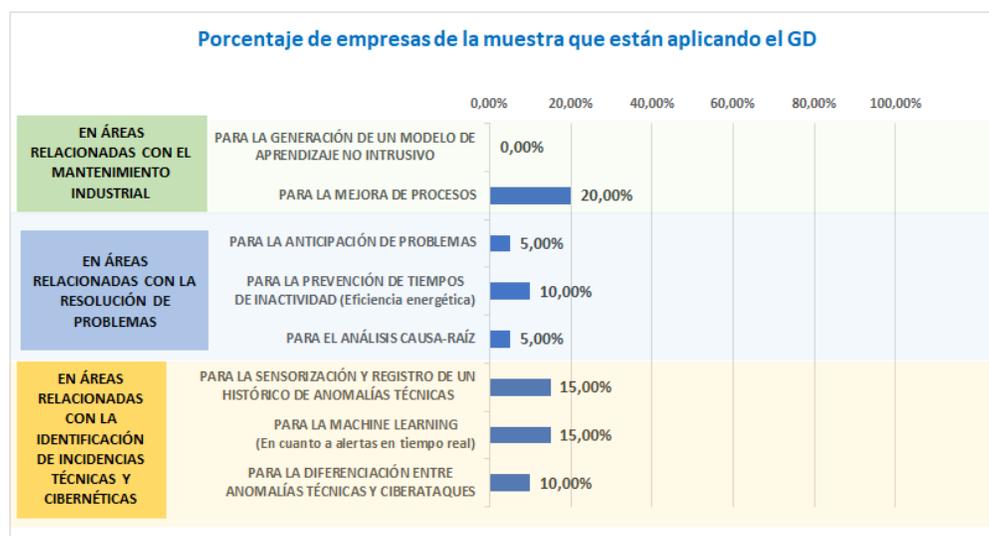

Figure 2: Autodiagnóstico sector Automoción: aplicación del gemelo digital

En la Figura 2 se muestra que solo una pequeña representación de empresas que han abordado algún proyecto relacionado esta tecnología con apoyo de la sensorización de los activos y el registro de anomalías (procesadas con técnicas de machine learning) lo aplican para la mejora de los procesos productivos y la identificación de incidencias técnicas. La implantación de metodologías de mantenimiento predictivo se enfrenta habitualmente a problemas de adopción de las nuevas tecnologías digitales, especialmente en sistemas de fabricación tradicional que suelen ser mayoritarios en PYMEs [Doyle and Cosgrove, 2019]. Por otro lado, es común encontrar maquinaria antigua o no digitalizada, y dónde existe digitalización los sistemas industriales comerciales utilizan fuentes de datos con acceso propietario e interfaces de comunicación heterogéneas para las que se desconoce la función (no abierta, con costes adicionales). De esta forma, la investigación de frameworks basados en el concepto de gemelo digital orientados a la generación de modelos de comportamiento predictivos para mejora continua se ha convertido en un punto clave a desarrollar dentro de la industria para el mantenimiento y funcionamiento inteligente de los equipos y sistemas [Mi et al., 2021]. Por lo tanto, el planteamiento ofrecido a los sistemas ciberfísicos junto a tecnologías de diagnóstico no intrusivas como el gemelo digital, abre una vía para hacer frente a una de las principales barreras actuales de entrada a nivel de planta en tareas de mantenimiento industrial [Madni et al., 2019]. Además, en otro apartado del autodiagnóstico se observa que las PYMEs están cada vez más sensibilizadas en cuanto a la importancia de invertir recursos en ciberseguridad. Existen trabajadores mucho más concienciados y formados, y, al menos la mitad de las empresas, tienen identificados todos los requisitos de seguridad recomendados, llegando a un cumplimiento del 90% en algunos casos. De este modo, la investigación de nuevas metodologías de caracterización y representación físico-digital de procesos y servicios sobre entornos reales conectados, se presenta como una alternativa esperanzadora para detectar, extraer y modelar amenazas de forma proactiva [García et al., 2022a].

---

[2]https://www.facyl.es/

## 2.2 Ciberseguridad en entornos convergentes

Las necesidades de la convergencia físico-digital han dado como resultado nuevos escenarios de ciberseguridad para la interconexión de los sistemas de control que gobiernan los ciclos de operación (SCADAs, PLCs, sensores y dispositivos de campo), con otras redes externas [Colbert and Kott, 2016]. Casi el 90% de las organizaciones están conectadas permanentemente a Internet, con servicios TI principalmente relacionados con proveedores de acceso de red de área amplia (WAN). Los accesos remotos a sistemas industriales se han generalizado, aumentando la exposición a riesgos y amenazas de ciberseguridad que no han sido evaluados convenientemente, y para los que las empresas suelen no están preparadas [Corallo et al., 2020]. Tradicionalmente, las redes industriales presentes en la mayoría de las plantas y sistemas industriales habían permanecido aislados de otros entornos [Semenkov et al., 2021]. Como resultado se presentan nuevas funcionalidades o servicios de monitorización remota o almacenamiento analítico de datos que provocan un aumento sustancial de los riesgos de seguridad en sistemas industriales, basado en nuevas amenazas específicas. Entre ellas, el aumento de vulnerabilidades asociadas a los ataques de ransomware contra infraestructuras industriales y el aumento del acceso remoto en un contexto de pandemia para explotar sistemas de control en tiempo real [Quintero-Bonilla and Martín del Rey, 2020]. Por lo tanto, se presentan nuevos desafíos para las empresas y los gestores de las mismas, que en ocasiones están limitados por las prácticas y restricciones de seguridad heredadas en entornos industriales [Rubio et al., 2019].

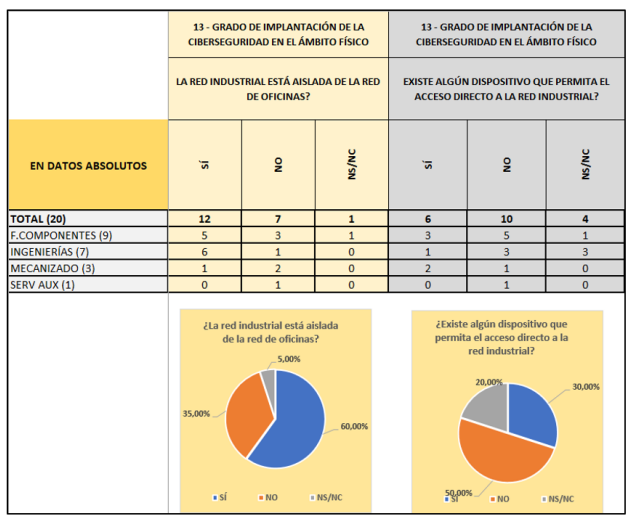

Figure 3: Autodiagnóstico sector Automoción: ciberseguridad

En la Figura 3 se analiza el grado de implantación de la ciberseguridad en el ámbito físico de acuerdo al autodiagnóstico realizado dentro del proyecto DETECTA a 20 empresas del sector de automoción. Se puede observar que todavía hay al menos un 35% de los casos en los que la red industrial y la red empresarial no están separadas, teniendo en cuenta la seguridad. Además, la digitalización introduce nuevos escenarios que requieren un acceso directo a la red industrial desde el exterior, lo que se manifiesta en un 30% de los casos. De forma complementaria se ha realizado otro diagnóstico entre los socios de la AEI de Ciberseguridad y Tecnologías Avanzadas[3], como proveedores y expertos en soluciones relativas a tecnologías habilitadoras de conocimiento predictivo y seguridad industrial. La muestra incluye 16 empresas de las que un 87,5% corresponden al segmento de PYMEs. Uno de los focos analizados ha sido el vector de entrada de las amenazas que supone mayor peligro en los ecosistemas industriales (ver Figura 4). De acuerdo a la concentración de respuestas, existen 3 focos principales que destacan sobre el resto: (i) Personas y procesos, (ii) Redes no segmentadas, y (iii) Aplicaciones no actualizadas.

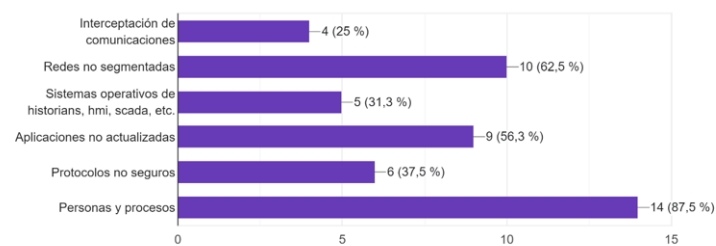

Figure 4: Diagnóstico sector Tecnologías Avanzadas: origen de amenazas de ciberseguridad

Respecto al peligro potencial de amenazas y su alcance, específicamente en un entorno industrial (ver Figura 5), destacan 4 respuestas sobre el resto: (i) Ramsonware, (ii) Malware o software malicioso, (iii) Vulnerabilidades de seguridad, y (iv) espionaje industrial.

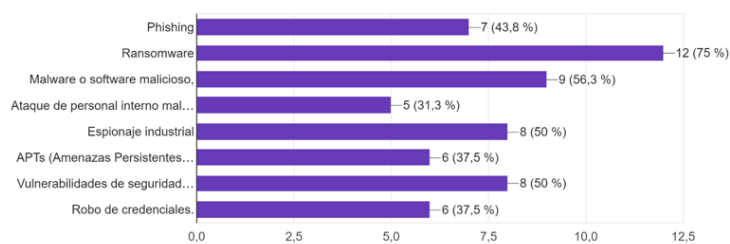

Figure 5: Diagnóstico sector Tecnologías Avanzadas: potencial de amenazas de ciberseguridad

Por lo tanto se evidencia la necesidad de nuevas estrategias de seguridad para las crecientes amenazas; por un lado, evaluando soluciones alternativas para hacer frente a los vectores de ataque industrial y, por otro lado, aumentando la rapidez y eficiencia en el posterior desarrollo de metodologías de prevención, defensa y respuesta [Conti et al., 2021]. En ese sentido, la caracterización físico-digital representa un enfoque aplicado para desarrollar métodos de ciberdefensa basados en el estudio y la clasificación de anomalías. Las nuevas tecnologías como los gemelos digitales de infraestructuras compuestas por sistemas de instrumentación y control, permiten probar la resistencia a ciberataques con alta fidelidad de forma no intrusiva y sin riesgo de daños reales al sistema [Alcaraz and Lopez, 2022]. Por otro lado, los modelos de detección y respuesta frente a ciberamenazas [Karantzas and Patsakis, 2021] permiten su aplicación para la mejora de la seguridad industrial con el apoyo de diferentes tecnologías, como la inteligencia artificial y el análisis de datos avanzados. Además, es posible construir modelos de aprendizaje de caja negra con la información recopilada de las características de funcionamiento del entorno de producción: (i) métricas de proceso, (ii) indicadores físicos, y (iii) patrones de operación monitorizados, junto a la utilización de soluciones basadas en la modelización de procesos para la identificación de anomalías [García et al., 2022b].

---

[3]https://www.aeiciberseguridad.es/

# 3 Metodología

El proyecto DETECTA introduce una visión orientada a la investigación de técnicas de aprendizaje no intrusivas para el diagnóstico predictivo, tanto desde el punto de vista de la mantenibilidad como de la ciberseguridad. De esta forma, afronta los problemas asociados a los diferentes estados de adopción de la tecnología en procesos industriales en las pymes, entendiendo aspectos clave donde intervienen frecuentemente procedimientos que dependen de máquinas y personas. Mediante la utilización de un entorno de laboratorio controlado y la experiencia de los trabajadores a pie de planta, profundiza en el conocimiento necesario que permita validar su aplicación futura sobre indicadores de mejora continua en entornos industriales. Por lo tanto, sin alterar el transcurso de la producción y sin que esto suponga un coste económico adicional, plantea la hibridación de diferentes tecnologías; por un lado con la virtualización de entornos de prueba de las infraestructuras industriales basados en gemelos digitales, y por otro con la aplicación de modelos de aprendizaje semi-supervisado. El objetivo es reproducir un entorno que permita:

- Caracterizar y explicar un conjunto de eventos técnicos dentro asociados a un entorno de fabricación industrial donde intervienen frecuentemente procedimientos que dependen de personas y máquinas (ciberseguridad y fiabilidad).
- Utilizar medios cognitivos para mejorar la toma de decisiones, basadas en algoritmos de aprendizaje que permitan apoyar acciones predictivas y la detección automática de anomalías para mejorar la ciberseguridad e incrementar la productividad de los procesos a pie de planta.

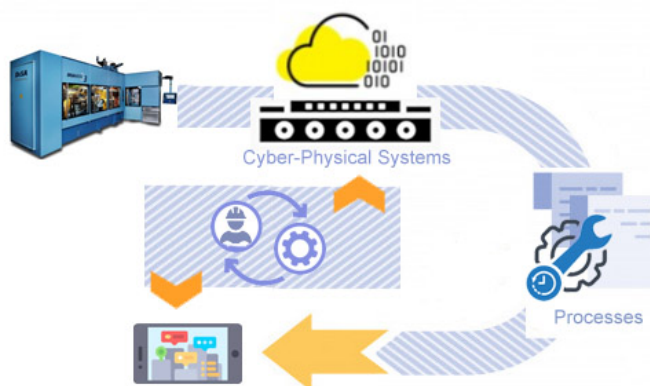

Figure 6: Ejemplo de ecosistema de interacción hombre-máquina en fabricación

De acuerdo a la Figura 6, la investigación de metodologías de generación de conocimiento no intrusivo para la fábrica digital y conectada presenta un nuevo paradigma ciberfísico, donde intervienen personas y máquinas, que une el ecosistema de la planta productiva y las nuevas tecnologías digitales. Para ello se han abordado tres fases diferenciadas que se describen a continuación.

**Fase 1. Análisis de los sistemas y personas que trabajan en el entorno productivo**.

El objetivo principal de esta primera fase es la identificación de perfiles clave en los departamentos implicados en el proceso de fabricación, así como el análisis y representación del funcionamiento de los sistemas industriales. Para ello, aborda el estudio de procedimientos a pie de planta, la forma de trabajo y el registro de datos que determinan el funcionamiento adecuado o anomalías durante el proceso (órdenes de trabajo, paradas, averías, mantenimiento, etc.). Como resultado se obtiene una recopilación de datos experimentales e indicadores procedentes del ecosistema de fabricación, incluyendo datos de operación y diagnóstico, datos de trabajo diarios, clasificación de las operaciones realizadas, herramientas y materiales utilizados, además de la interacción con la cadena de suministro y soporte técnico remoto.

**Fase 2. Pruebas de laboratorio**.

Durante la segunda fase, la visión representada por el gemelo digital permite la generación de ecosistemas de aprendizaje para la toma de decisiones de forma más rápida y ágil: (i) teniendo información del ciclo de vida de los diferentes procesos y sus interacciones con la posibilidad de generar conocimiento a partir de sus estados e indicadores en tiempo real ; (ii) analizando decisiones estratégicas de forma centralizada y efectiva sin interferir o paralizar los procesos productivos, anticipándose con un enfoque predictivo, no reactivo , y (iii) incorporando modelos de datos distribuidos a través de diferentes puntos de la cadena de proceso y de suministro. La caracterización de estos modelos de gemelo digital y caja negra, de forma no intrusiva, utiliza conjuntos de datos (datasets) reales y etiquetados del entorno industrial para definir patrones o huellas personalizados que se puedan asociar a un comportamiento de un sistema/proceso durante periodos de su ciclo de vida. La investigación y creación de modelos de inteligencia artificial, de forma no intrusiva, utiliza como base una interfaz con los datos y los sistemas caracterizados para la generación de patrones de conocimiento apoyados por los expertos de proceso, de forma que sea posible abordar diferentes formas de detección de eventos técnicos de forma híbrida.

**Fase 3. Validación con entornos de aprendizaje basados en el concepto de gemelo digital**

Finalmente, la última fase permite la validación del aprendizaje adquirido de forma predictiva. En primer lugar, a través de pruebas donde se validan los patrones de conocimiento de los procesos para detección de anomalías (predicción de averías, paradas por mantenimiento y ciberataques), donde se utilizan diferentes estrategias: (i) exploración con modelos de aprendizaje automático, (ii) generación de métricas correspondientes a las técnicas de aprendizaje automático seleccionadas, y (iii) generación de pruebas de caja negra a nivel de especificaciones de modelos de comportamiento de eventos discretos. En segundo lugar, mediante la validación por conocimiento experto de los criterios anómalos, donde se emplean los puntos de control e indicadores para medir el impacto sobre el proceso de mecanizado y los riesgos que se han utilizado para medir el impacto sobre dichos procesos. El resultado final es una matriz de resultados de acuerdo a las causas y riesgos asociados.

## 4 Creación de componentes

### 4.1 Interfaz virtual con indicadores de operación y diagnóstico

De acuerdo a la metodología presentada, el primer componente corresponde a una representación virtual de variables e indicadores físicos del escenario de fabricación industrial investigado. En este escenario intervienen tanto trabajadores como máquinas, por lo tanto presenta la funcionalidad del concepto del gemelo digital para la visualización y parametrización de los valores diarios obtenidos y analizados durante el funcionamiento del sistema (ver Figura 7).

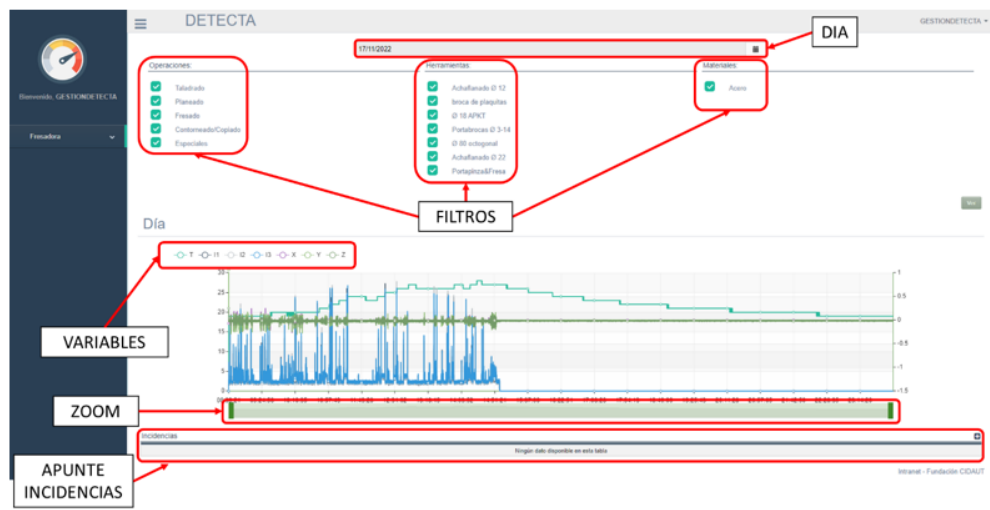

Figure 7: Visualización de datos del gemelo digital del sistema de mecanizado industrial

Este enfoque permite la caracterización del proceso de mecanizado industrial y la visualización de indicadores relativos a las variables monitorizadas (consumo, vibraciones, temperatura, etc.), a la vez que se investiga la detección de anomalías e incidentes de ciberseguridad de forma no intrusiva con dicha información. Para registrar la intensidad de corriente trifásica que alimenta a la fresadora se cargan datos de tres sensores de lazo cerrado (bobina Rogowski), mientras que la lectura de datos de temperatura se obtiene en el cabezal de mecanizado a través de una sonda no intrusiva RTD (detector de temperatura por resistencia). Por otro lado, la lectura de datos de aceleración se realiza mediante un acelerómetro de tres ejes (X, Y, Z) también en el cabezal de mecanizado. La interfaz virtual además incorpora herramientas para el modelado híbrido de anomalías con inteligencia artificial y sistemas de caja negra. De esta forma cuenta con diferentes funcionalidades para incrementar el conocimiento en las operaciones y el diagnóstico. Por ejemplo, permite elegir un día en concreto y visualizar cada una de las variables registradas de manera simultánea e independiente. También es posible filtrar los datos a visualizar por cada operación realizada por el trabajador (taladrado, fresado, mandrinado, etc.), herramienta utilizada y material manipulado. Esta opción permite identificar cómo se comporta el sistema de mecanizado en base a patrones específicos.

## 4.2 Modelo de detección de alerta temprana

En el campo de la detección de alerta temprana, la capacidad de predecir eventos, acciones o comportamientos adquiere una gran importancia en diversos ámbitos y casos de uso. La expansión de la inteligencia artificial y, en particular, el avance de las técnicas de machine learning, ha permitido desarrollar modelos capaces de realizar predicciones con anticipación y precisión. El presente proyecto se enfoca en la predicción de eventos y acciones relacionados con la ciberseguridad, así como en la detección de posibles averías o incidencias que puedan afectar la productividad. Para lograr este objetivo, se emplean técnicas de series temporales, las cuales son ampliamente utilizadas en la investigación de procesos de producción. En el ámbito de las series temporales, se han utilizado métodos de detección de anomalías para la identificación de irregularidades en los procesos de fabricación [Hsieh et al., 2019]. Además, estos métodos también se aplican en la detección de fallos en procesos industriales, contribuyendo al mantenimiento predictivo [Chadha et al., 2019].

Asimismo, el mantenimiento de procesos y maquinaria industrial puede beneficiarse de la capacidad predictiva. En este sentido, el trabajo realizado por Kanawaday et al. destaca por el uso de modelos ARIMA en la predicción de posibles errores y defectos en la calidad de la producción [Kanawaday and Sane, 2017]. En este proyecto en particular, se plantea la predicción de anomalías con el objetivo de detectar posibles malfuncionamientos o ciberataques en el proceso de producción. Además, se busca predecir acciones de mantenimiento o errores recurrentes en el tiempo. Para ello, se utilizarán técnicas predictivas en series temporales, las cuales permiten adaptarse a los procesos de producción dinámicos. En caso de ser necesario, se realizará un diseño ad hoc de técnicas de Machine Learning para satisfacer las necesidades específicas del proceso industrial. En un trabajo preliminar con un enfoque similar, Kapp et al. proponen un sistema de Automated Machine Learning que combina varias técnicas y permite una monitorización avanzada para predecir posibles errores y evitar interrupciones completas en la producción [Kapp et al., 2020].

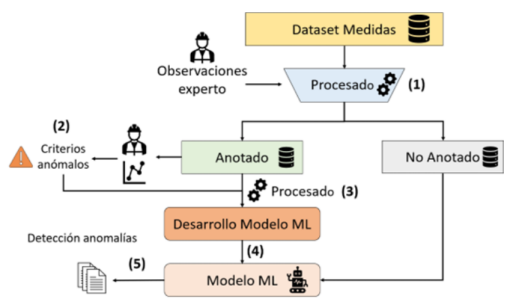

Figure 8: Modelo de detección de alerta temprana

Tal y como muestra la Figura 8, el modelo de detección de alerta temprana se basa en una metodología que involucra varios pasos. En primer lugar, se utiliza un conjunto de datos en bruto que contiene medidas de sensores de temperatura, vibraciones y corriente durante un período determinado. Además, se cuenta con observaciones realizadas por un experto en un subconjunto de estos datos, lo que permite diferenciar entre datos anotados y no anotados. En el segundo paso, se colabora con el experto para extraer y validar criterios automáticos basados en reglas para la detección de anomalías. Estos criterios se aplican sobre los datos anotados, lo que da como resultado un conjunto de datos etiquetados. A continuación, se realiza un análisis exploratorio de los criterios en los datos, identificando diferentes tipos de anomalías y agrupándolos en clases de anomalías. Una vez definidas las clases de anomalías, se procede al desarrollo, evaluación y selección de diversos modelos de Aprendizaje Automático. Mediante este proceso, se elige el mejor modelo utilizando los datos con observaciones. Finalmente, se utiliza el modelo seleccionado para detectar anomalías en el conjunto total de datos iniciales. El resultado es la generación de un archivo con formato separado por comas (CSV) que extiende la detección de anomalías a todo el conjunto de datos de medidas, permitiendo la predicción de anomalías en un marco temporal de 30 segundos.

### 4.3 Modelo de detección basado en caja negra

Uno de los retos abordados en este proyecto ha sido la caracterización sin limitar la productividad y la disponibilidad permanente de información. A este respecto, la investigación de la modelización de procesos en entornos de fabricación industrial mediante Técnicas de Diseño y Análisis Estructurado (SADT) [Marca and McGowan, 1987] apoya el enfoque híbrido aplicado a tareas de mantenimiento ciberseguro a través de la detección de anomalías. SADT permite estructurar un problema definiendo todos sus componentes (entidades, recursos y controles) en un nivel inferior de modelado. Esto ayuda a identificar las interacciones individuales entre estos componentes junto con cualquier recurso o requisito. Las notaciones SADT consisten en diagramas de caja-flecha (bloques), con cuatro flechas en cada lado definidas como: entrada, salida, control y una actividad en el medio como se muestra en la Figura 9:

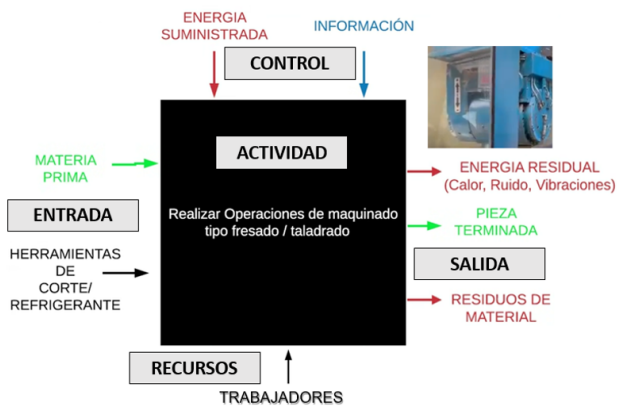

Figure 9: Ejemplo de aplicación de un modelo SADT sobre el proceso de mecanizado

- Actividad: Una actividad es cualquier función o proceso que sirve para transformar insumos en productos.
- Entrada: Los datos/información requerida por una actividad para iniciar el proceso de transformación.
- Salida: los datos/información producidos por la actividad como resultado de esta transformación.
- Control: Cualquier restricción que afecta el comportamiento de la actividad de alguna manera.
- Mecanismo: Personas, recursos o cualquier medio que se requiere para llevar a cabo la actividad.

El objetivo buscado con este componente es disponer de un entorno experimental donde abordar casos de uso relativos a la ciberseguridad y realizar un análisis causa-raíz utilizando metodologías de caja negra. Se busca una caracterización de acuerdo a un patrón o relación matemática que sea capaz de predecir el comportamiento de las variables de interés. En este escenario se ha considerado el principal vector de ataque el originado en situaciones de acceso remoto. Por lo tanto, el centro de mecanizado se plantea conectado a través de una red WiFi o cableada que da acceso a un router que tiene una dirección IP y un puerto accesible desde Internet para labores de mantenimiento por una empresa externa. La existencia de vulnerabilidades asociadas a los sistemas informáticos abre la puerta a una potencial brecha de seguridad. Por ejemplo, a través de este acceso un atacante podría alcanzar el acceso al equipo hardware que aloja el sistema de control de la fresadora y realizar cambios en el programa o actuar sobre la propia máquina.

## 5 Resultados

Durante la ejecución del proyecto DETECTA se ha generado conocimiento para la mejora del mantenimiento industrial en procesos de fabricación, manteniendo la productividad de los medios de fabricación a la vez que se garantiza la ciberseguridad. Los beneficios que representa son muy importantes, incorporando métodos de caracterización no intrusivos para la detección de anomalías y reducción de riesgos asociados a tareas de asistencia remota. Como análisis de la aplicabilidad de las metodologías y tecnologías presentadas durante el proyecto, se incluye una matriz de resultados en la que han participado para su evaluación los diferentes socios, destacando la participación de la PYME industrial del sector de automoción.

En los siguientes apartados se describen los resultados alcanzados a partir de la evaluación de tres objetivos principales:

- Entender y caracterizar el funcionamiento de un sistema de mecanizado en un entorno productivo.

- Detectar e identificar posibles problemas relativos al mantenimiento del sistema industrial.
- Detectar e identificar posibles riesgos de ciberataque al sistema industrial.

### 5.1 Modelos de datos del gemelo digital

El proceso elegido para investigar los sistemas de producción es muy representativo en PYMEs que trabajan en series de fabricación de componentes homogéneos y heterogéneos indistintamente. En concreto, se ha caracterizado un sistema de fabricación industrial por mecanizado ubicado en Industrias Maxi que integra los componentes en procesos auxiliares del sector de automoción. Para la generación de conocimiento aplicado en tareas de mantenimiento predictivo y detección de incidentes de ciberseguridad se ha diseñado una interfaz de gemelo digital con indicadores de operación y diagnóstico, como se muestra en la Figura 10.

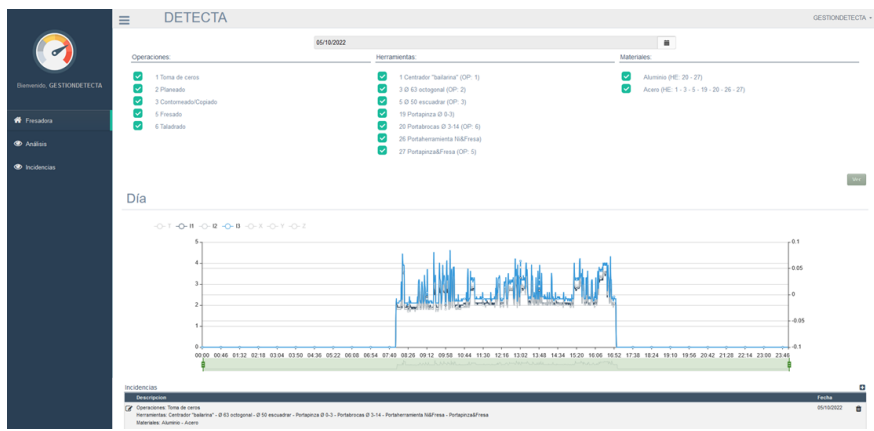

Figure 10: Interfaz virtual y modelo de datos del gemelo digital del sistema de frabricación por mecanizado

La interfaz permite representar y parametrizar el funcionamiento del sistema de mecanizado con la incorporación de modelos de machine learning y caja negra, a partir de la recopilación del conjunto de datos experimentales e indicadores procedentes del proceso de mecanizado:

- OPERACIONES: Taladrado, Planeado, Fresado, Contorneado/Copiado y Especiales.
- HERRAMIENTAS: Achaflanado ø12, ø18 APKT, Portabrocas 3-14, ø50 escuadrar, Achaflanado ø22, ø63 octogonal, Portapinzas&Fresa, etc.
- MATERIALES: Acero, Aluminio, Plástico, Acero Inox, etc.
- TIPO DE ACCESO: Local, Remoto.
- VARIABLES: Temperatura, intensidades de corriente trifásica, aceleración en tres ejes y registro de incidencias anotadas por los trabajadores implicados en el proceso.

Se han seleccionado datos entre los meses de agosto y noviembre de 2022. Para cada día aparecen las opciones disponibles asociadas para optimizar el tiempo de observación y facilitar el análisis de los datos. De ese modo es posible clasificar de manera precisa la funcionalidad que se va a estudiar con todos los datos disponibles. Por último incluye un registro de incidencias etiquetadas por un trabajador experto, lo que facilita la introducción de información acerca del proceso realizado en un periodo concreto. De este modo, la información anotada sobre un suceso relevante, queda guardado junto con los datos registrados de su serie temporal. En la Figura 10 se pueden observar las operaciones, herramientas, materiales, variables de proceso e incidencias registradas para el día 5 de octubre de 2022. Toda esta información, almacenada en una base de datos de series temporales, se utiliza para la visualización y el análisis de modelos de registro de incidentes (mantenimiento y ciberseguridad) con la ayuda del responsable de fabricación e IT, operadores de máquina, técnicos de mantenimiento y el responsable de mejora continua.

### 5.2 Detección de anomalías

Para el desarrollo del sistema inteligente encargado de la detección y predicción de anomalías se ha contado con los datos en bruto registrados por el gemelo digital y la supervisión de un trabajador experto de Industrias Maxi. Incluyen medidas de sensores de temperatura, consumo de corriente, vibración detectada por acelerómetros asociados a la maquinaria, así como el tipo de operación, material y herramienta empleada a cada segundo durante los meses indicados en el apartado anterior. De este modo, estos datos incluyen 7.028.482 de medidas correspondientes a un total de 88 días sobre los que se han analizado los modelos seleccionados, las métricas de evaluación y la visualización de los árboles de decisión que conforman los modelos.

Específicamente, durante el proceso de detección de anomalías se han utilizado dos enfoques: aproximación binaria y aproximación multi-clase. En la aproximación binaria, se han utilizado ventanas temporales de 15 minutos. Se compararon varios modelos mediante validación cruzada de 10-folds, y se obtuvo que el modelo de Random Forest alcanzó las mejores métricas de evaluación. Uno de los aspectos destacables es que el modelo logró identificar valores de corriente máxima que concuerdan con el límite establecido por el experto, lo que respalda el potencial del aprendizaje automático para detectar anomalías. Además, en la visualización de uno de los árboles individuales que conforman el conjunto del modelo Random Forest presentado en la Figura 11, se muestran como resultado los umbrales de corriente máxima utilizados para la detección de anomalías.

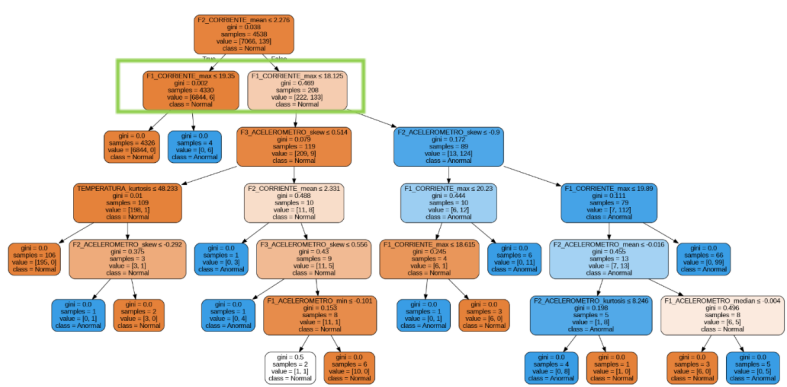

Figure 11: Visualización de árboles de decisión que forman parte del ensemble de Random Forest binario en ventanas de 15 minutos. Se pueden observar las variables y umbrales empleadas por los modelos para detectar anomalías. En verde se señalan las hojas del árbol que muestran los umbrales de corriente máxima establecidos por el modelo que se encuentran en línea con el límite establecido por el experto.

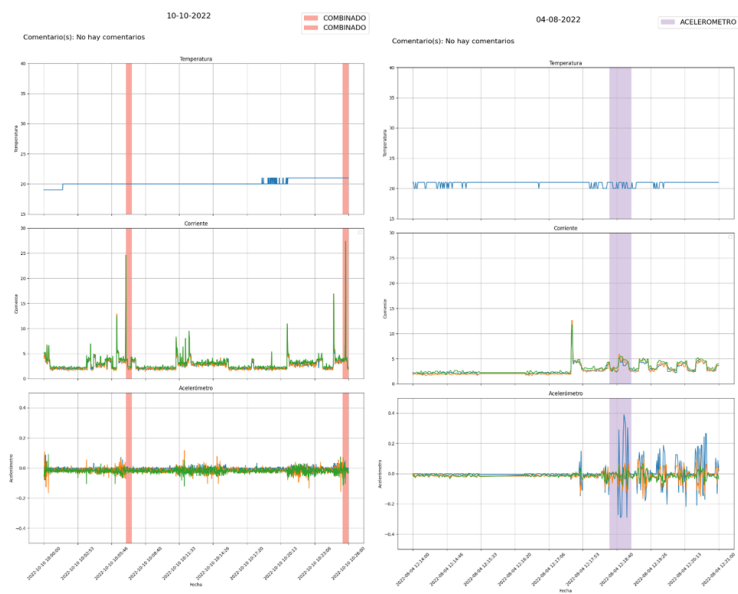

Figure 12: Visualización de las ventanas temporales de 30 segundos detectadas por el motor de aprendizaje automático multi-clase. En la leyenda se indican los distintos tipos de anomalías detectados

En la aproximación multi-clase, se utilizaron ventanas temporales de 30 segundos. Los tres mejores modelos seleccionados fueron Random Forest Classifier, Decision Tree Classifier y Extra Trees Classifier, debido a su alta precisión, recall y F1-macro. En la Figura 12 se muestran las medidas de los sensores de temperatura, corriente y acelerómetro para diferentes días de uso de trabajo de la máquina. Esta figura destaca las ventanas temporales de 30 segundos detectadas como anómalas por el modelo, las cuales se ajustaron a los criterios establecidos por el experto. En cuanto al rendimiento de los modelos de aprendizaje automático, se analizó el tiempo de cómputo empleado para el entrenamiento y la inferencia. Se observó que el tiempo de entrenamiento aumenta a medida que se reducen las ventanas temporales, siendo la aproximación binaria con ventanas de 15 minutos la que requirió menor tiempo. Por otro lado, el tiempo de inferencia para el modelo multi-clase se estimó en 6.94 ± 1.97 segundos para el conjunto completo de ventanas temporales de 30 segundos.

En resumen, los resultados obtenidos en la detección de anomalías enfocadas a alerta temprana demostraron la eficacia de los modelos de aprendizaje automático, tanto en la aproximación binaria como en la multi-clase. Estos modelos lograron identificar patrones anómalos en las medidas de los sensores, brindando una alta precisión en la detección de anomalías y proporcionando información relevante para la toma de decisiones. Para la generación de la matriz final descrita en la siguiente sección, se tuvieron en cuenta las diferentes casuísticas identificadas en la planta de Industrias Maxi de acuerdo a fallos del sistema de fabricación y a potenciales incidencias de ciberseguridad como se muestra en la Figura 13.

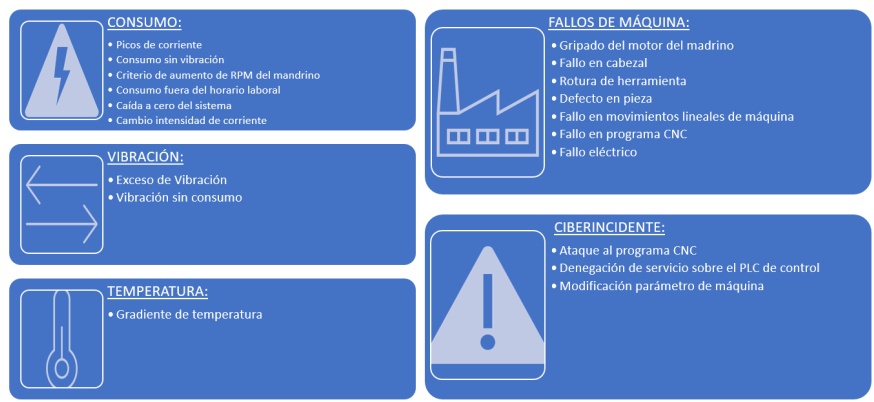

Figure 13: Casuísticas identificadas de acuerdo a anomalías por fallo de máquina o por ciberincidente

Como ejemplo de anomalías detectadas tenemos el aumento de régimen de revoluciones en el sistema de mecanizado. Esta anomalía conduce a un posible gripado con afectación a la herramienta y a la pieza mecanizada. En la Figura 14 se observa una operación que presenta cambios en la intensidad de corriente para un tipo de operación concreto y consumos muy elevados de forma continuada acompañada de una subida de temperatura importante. Además, se observa que pasados unos minutos la máquina se para completamente.

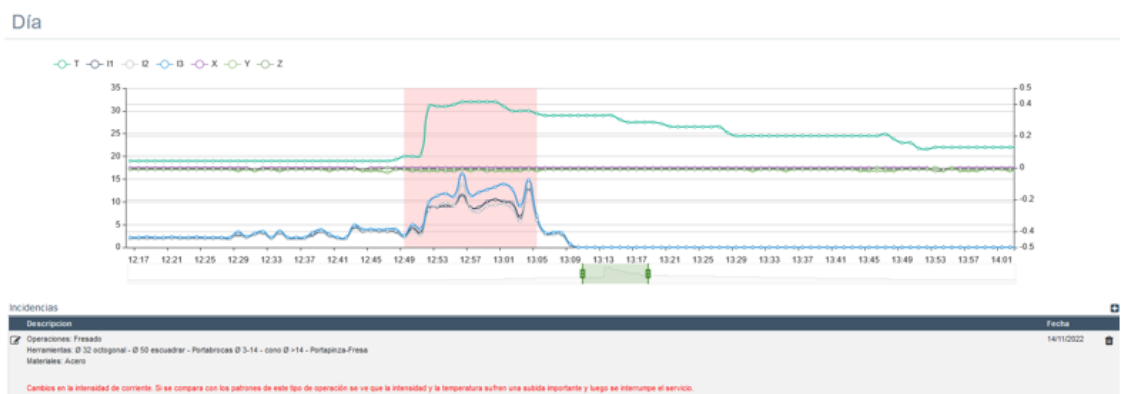

Figure 14: Ejemplo de anomalía detectada por aumento de régimen de revoluciones en el mandrino

## 5.3 Matriz causa/riesgo

Como se ha presentado en los apartados anteriores, la ejecución del proyecto DETECTA va a ayudar a las PYMES industriales a encontrar los caminos que permitan mitigar los riesgos de mantenibilidad y ciberseguridad. Durante el transcurso del proyecto se han integrado funcionalidades basadas en las tecnologías digitales emergentes. Esto ha permitido el desarrollo de soluciones no intrusivas que hagan más robustos los medios de producción de un sistema de mecanizado industrial (Figura 15). Entre ellas:

1. Aplicación del concepto de gemelo digital para la caracterización de procesos y detección híbrida de anomalías como palanca predictiva en entornos convergentes.
2. Motor de inteligencia para detección temprana de anomalías y mantenimiento predictivo y ciberseguro de sistemas y componentes de fabricación mediante algoritmos de aprendizaje automático.

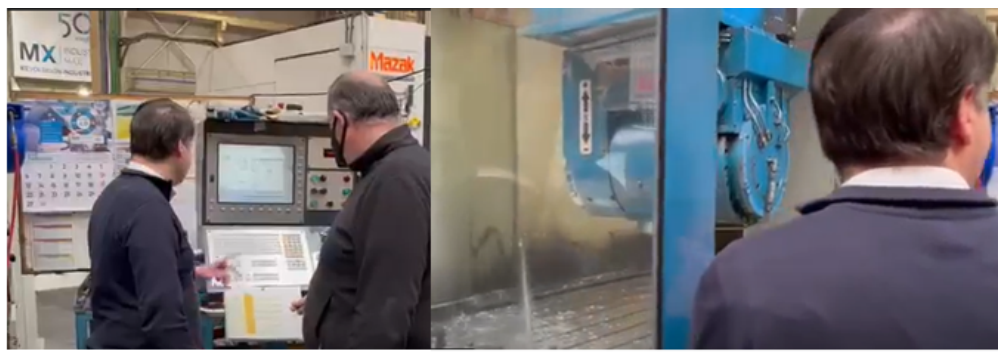

Figure 15: Centro de mecanizado para fabricación industrial ubicado en Industrias Maxi

Por tanto, se ha podido abordar:

- El procesamiento y extracción de datos útiles para la formalización de criterios de detección de anomalías, validados en colaboración con expertos en la materia.
- El empleo de estos criterios para la generación de un dataset etiquetado que permita el desarrollo de modelos supervisados de aprendizaje automático.
- El desarrollo y evaluación de modelos de aprendizaje automático para la detección y clasificación de anomalías.

El resultado final es una matriz de análisis detallada en la Figura 16 con anomalías y riesgos originados sobre un escenario industrial conectado y altamente competitivo.

Figure 16: Matriz de resultados de acuerdo a las causas y riesgos asociados a los procesos analizados

# 6 Conclusiones y trabajo futuro

La monitorización y detección de anomalías de forma predictiva en entornos de fabricación de componentes industriales tiene como objetivo anticiparse a las incidencias ocasionadas habitualmente por roturas de componentes o por errores de funcionamiento de los sistemas. Muchas de las máquinas industriales se consideraban elementos aislados con bajos niveles de seguridad y estrategias tradicionales de mantenimiento; ahora, con los nuevos requisitos de productividad y digitalización de las cadenas de suministro en las PYMEs, la convergencia físico-digital introduce potenciales riesgos asociados a la conectividad de los sistemas. A pesar del desarrollo de tecnologías de ciberseguridad, su aplicación en entornos de operación industrial altamente heterogéneos está limitada por carencias en las políticas de seguridad heredadas. Esto influye, tanto en las soluciones de reingeniería para contemplar diferentes tipologías y vectores de ataque industrial, como en la rapidez y eficacia para el desarrollo de metodologías de prevención, defensa y respuesta frente a las nuevas amenazas. Se concentran, por tanto, diferentes anomalías que pueden afectar al proceso de producción, existiendo casos muy complejos que necesitan primero de un estudio de posibles incidencias que pueden sufrir las máquinas y por otro lado las medidas de seguridad, detección y respuesta adecuadas.

Con la colaboración de centros de conocimiento y expertos en procesos industriales, el proyecto DETECTA ha permitido la adopción de un modelo de conocimiento híbrido basado en la caracterización y detección de anomalías. El proyecto responde así a varios retos tecnológicos planteados desde el punto de vista de la mantenibilidad y ciberseguridad en diferentes fases: (i) investigando técnicas para la caracterización de procesos y detección de anomalías de forma no intrusiva apoyadas tanto en el concepto de gemelo digital, como en algoritmos de inteligencia artificial, (ii) analizando procesos de fabricación industrial en un ecosistema PYME donde entender los indicadores asociados a la disponibilidad de los sistemas y las medidas necesarias para su protección frente a riesgos y amenazas, (iii) modelando entornos de aprendizaje en sistemas de caja negra que permitan discriminar la predicción de averías, paradas por mantenimiento y ciberataques, (iv) analizando los retos que suponen los entornos convergentes de Tecnologías de Operación y Tecnologías de la Información que exponen medios productivos a Internet, y (v) evaluando un caso industrial en una PYMEs del sector de fabricación para la generación de una matriz de resultados a partir de la integración de los diferentes componentes tecnológicos obtenidos mediante la investigación planificada y pruebas de laboratorio.

A partir de los componentes implementados se plantea una segunda fase de investigación para la construcción de prototipos de escala relevante con nuevos sensores de mayor precisión. El objetivo es extender las capacidades de inteligencia artificial que permitan una mayor precisión en la detección de anomalías y su activación de forma proactiva, a la vez que se validan en el entorno de la PYME industrial.